\documentclass[12pt]{iopart}
\usepackage{graphicx}

\newcommand{\D}{\ensuremath{\mathrm{d}}}
\renewcommand{\Re}{\ensuremath{\mathrm{Re}}}

\newcommand{\Cee}{\ensuremath{\mathcal{C}}}

\def\-{{\bf --}} 

\def\(#1){(\ref{#1})}

\begin{document}


\title{Dyck Paths, Motzkin Paths and Traffic Jams}
\date{May 2004; Revised July 2004}

\author{R.\ A.\ Blythe}
\address{Department of Physics and Astronomy, University of Manchester,
Manchester, M13 9PL, England}

\author{W.\  Janke}
\address{Institut f\"ur Theoretische Physik,
 Universit\"at Leipzig,
Augustusplatz 10/11,
04109 Leipzig, Germany}

\author{D.\ A.\ Johnston}
\address{School of Mathematical and Computer Sciences,
Heriot-Watt University, Riccarton, Edinburgh EH14 4AS, Scotland}

\author{R.\ Kenna}
\address{School of Mathematical and Information Sciences,
Coventry University,
Coventry, CV1 5FB, England}

\begin{abstract}
It has recently been observed that the normalisation of a
one-dimensional out-of-equilibrium model, the Asymmetric Exclusion
Process (ASEP) with random sequential dynamics, is exactly equivalent
to the partition function of a two-dimensional lattice path model of
one-transit walks, or equivalently Dyck paths.  This explains the
applicability of the Lee-Yang theory of partition function zeros to
the ASEP normalisation.
 
In this paper we consider the exact solution of the {\it
parallel\/}-update ASEP, a special case of the Nagel-Schreckenberg
model for traffic flow, in which the ASEP phase transitions can be
intepreted as jamming transitions, and find that Lee-Yang theory still
applies. We show that the parallel-update ASEP normalisation can be
expressed as one of several equivalent two-dimensional lattice path
problems involving weighted Dyck or Motzkin paths.  We introduce the
notion of thermodynamic equivalence for such paths and show that the
robustness of the general form of the ASEP phase diagram under various
update dynamics is a consequence of this thermodynamic equivalence.

\end{abstract}


\maketitle


\section{Introduction}

It is a fact of equilibrium statistical-mechanical life that if one
wishes to establish the thermodynamics of a model system it is
necessary to calculate its partition function.  When dealing with
\textit{nonequilibrium} stationary states (for example those that
carry a current) it is not obvious that the object analogous to the
equilibrium partition function, namely an appropriately defined
normalisation of the stationary distribution, encodes the
thermodynamics in a similar way.

By now the Lee-Yang zeros of such a normalisation for a number of
nonequilibrium steady states have been studied in the complex plane of
the microscopically irreversible transition rates that define a
particular model.  In particular, various one-dimensional driven
diffusive systems \cite{Arndt,BE1}, reaction-diffusion processes
\cite{DDH,Jafar1,Jafar2} and an urn model for the separation of sand
\cite{Droz} have been investigated and, in each case, the zeros
approach the real axis at the appropriate transition points (aspects
of this work have been reviewed in \cite{BEbjp}).  Furthermore, the
locus (i.e., impact angle) and density of the zeros near the
transition point correspond to the order of the phase transition
(defined in physical terms according to whether there is, e.g., phase
coexistence or a diverging correlation length) in exactly the same way
as for the equilibrium systems long ago considered by Lee and Yang
\cite{YL,LY}.

A striking feature of these various analyses is that the role of the
\textit{equilibrium} fugacities in the Lee-Yang approach is taken up
by the \textit{nonequilibrium} transition rates present in the model.
The aim of this work is to explore this connection more deeply, the
main result being that one can find equilibrium models with the same
phase behaviour as the original nonequilibrium model, but within which
the transition rates are transparently fugacities.

At a rather abstract level, this phenomenon can be understood in terms
of a graph-theoretic description of the microscopic process
\cite{RABthesis,Brak2}.  In this picture, one considers a graph in
which each vertex corresponds to a microscopic configuration and
directed edges to the transitions permitted in an elementary timestep.
A representation of the steady-state normalisation can then be shown
to coincide with the generating function of spanning in-trees on this
graph \cite{Harary}.  Explicitly, if the microscopic transition rates
(or, in a discrete-time process, transition probabilities) are drawn
from the set $\{ w_1, w_2, \ldots, w_N\}$, the normalisation $Z$ can
be expressed as
\begin{equation}
\label{trees}
Z = \sum_{m_1, \ldots, m_N} n(m_1, \ldots, m_N) w_1^{m_1} w_2^{m_2}
\cdots w_N^{m_N} \;,
\end{equation}
in which $n(m_1, \ldots, m_N)$ counts the number of spanning in-trees
on the graph of configurations with $m_i$ edges corresponding to a
transition that occurs with rate $w_i$.  Thus the rates $w_i$ act as
fugacities that control the amount to which various classes of
spanning in-tree contribute to the overall normalisation.  One can see
there is the possibility for different classes of trees to dominate
the normalisation as the fugacities that weight the trees are varied.
One may conjecture that such changes correspond to thermodynamic phase
transitions in the microscopic model.

The representation of the normalisation (\ref{trees}) is appealing in
that it is uniquely defined for any stochastic process with a single
steady state and polynomial in the variables $w_i$.  In principle,
this form can be obtained by summing the principal minors derived from
the matrix of transition rates, wherein each summand corresponds to
the steady-state weight of a particular configuration
\cite{RABthesis,EB}.  In practice, however, evaluation of the
determinants involved is intractable, and in the cases where
nonequilibrium steady-state distributions have been obtained via
specialised methods, one typically finds polynomials that are lower in
degree than (\ref{trees}).  We believe that in these cases, the two
distributions are related by a factor common to all the
configurational weights which does not develop any nonanalyticities in
the thermodynamic limit.  In this instance, a Lee-Yang analysis gives
the same results independently of the form of the normalisation
studied and therefore we do not worry unduly that the normalisation is
generally not uniquely defined.

A greater difficulty with expressions of the form (\ref{trees}) is
that the relationship between fugacities and physical observables is
unclear.  Recently, however, Brak \textit{et al.}~\cite{Brak2}
elucidated the connection for the asymmetric simple exclusion process
(ASEP) with open boundaries, a model that was solved exactly and
independently in \cite{DEHP} and \cite{ScDo}.  There, the (reduced)
normalisation of the ASEP was related to the partition function of an
equilibrium surface model in which the two transition rates associated
with the boundaries turn out to be fugacities conjugate to the
densities of surface contacts with the horizontal axis from above and
below.  Thus for this model, there is an interpretation of the
normalisation from which the thermodynamics can be extracted using
standard equilibrium statistical mechanical techniques.

This result raises the question of whether the relationship between
nonequilibrium transition rates and equilibrium fugacities extends to
other nonequilibrium steady states.  In this work we investigate this
question by revisiting another exactly-solved variant of the ASEP,
namely that with \textit{parallel} lattice updates \cite{TE,Ev1,dGN}.
Whereas the case with a random-sequential updating scheme, referred to
above, was introduced to model the kinetics of biopolymerisation
\cite{MGP}, the parallel version arises as a special case of the
Nagel-Schreckenberg model of traffic flow \cite{NS}.  From this point
of view, the phase transitions in the model's steady state can be
related to traffic jamming.

We begin by recalling the definition of the parallel-update ASEP along
with some key results.  In addition to the two boundary parameters of
the random-sequential variant, there is a further quantity, $p$, which
relates to the degree of parallelism in the dynamics.  This gives rise
to a normalisation with a considerably more complicated structure than
that for the random-sequential ASEP.  It is therefore appropriate to
check that the Lee-Yang zeros of this normalisation correctly
reproduce the known phase behaviour for the model in the complex plane
of the defining transition probabilities. This being the case, we move
on to establish partition functions for equilibrium surfaces or,
equivalently, random walks on two different lattices that have the
same thermodynamic phase behaviour as the parallel-update ASEP (and
thence jamming in its guise as a traffic model).  In each case the
extra parameter, $p$, generalises the surfaces or walks obtained for
the random-sequential ASEP in a physically meaningful way.  In one
case, $p$ is a fugacity related to the density of horizontal segments
of walks on a triangular lattice and in the other it is related to the
density of peaks on the surface.  In order to obtain these results we
find that it is of great benefit to consider a
\textit{grand-canonical} ensemble in which the system size is a
quantity exhibiting equilibrium fluctuations about its mean.  By
considering the general equilibrium thermodynamic properties of the
surfaces, we are able to gain a further insight into the robustness of
the phase diagram for the ASEP under different microscopic updating
schemes.

\section{The Parallel-Update ASEP}
\label{definition}

The microscopic dynamics of the asymmetric simple exclusion process
with parallel dynamics are as shown schematically in Fig.~1 and
defined as follows. Particles are introduced with probability $\alpha$
in each timestep at the start of a one-dimensional lattice with $L$
sites (if the first site is empty) and leave with probability $\beta$
at the other end. They can hop with probability $p$ one unit to the
right if that space is empty (or remain still with probability $1-p$),
otherwise if the space to the right is blocked they remain stationary.
The update is applied to all particles simultaneously in contrast to
the random sequential update scheme.  Nevertheless, the latter is
recovered in the limit $p \propto \D t \to 0$ (where $\D t$ is the
size of each timestep) if $\alpha$ and $\beta$ are rescaled to remain
proportional to $p$.

\begin{figure}
\begin{center}
\includegraphics[scale=0.6]{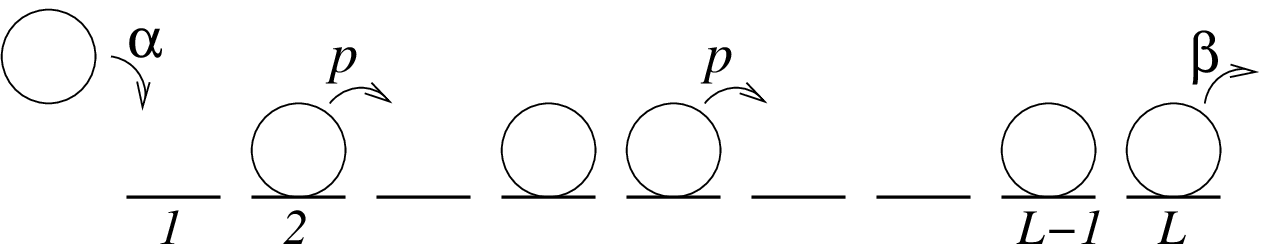}
\caption{\label{asep-fig}Dynamics of the ASEP.  The arrow labels
indicate the probabilities with which the corresponding transitions
occur; site labels are also indicated.  Note that the moves are
conducted in parallel.}
\end{center}
\end{figure}

Quantities such as the current and density correlation functions for
the ASEP with fully parallel dynamics ($p=1$) were first worked out by
Tilstra and Ernst \cite{TE}.  Although a cluster approximation was
used, the results obtained were believed to be correct in the
thermodynamic limit ($L\to\infty$).  This claim was subsequently
confirmed in the independent works of \cite{Ev1} and \cite{dGN} which
employed two different matrix product ans\"atze \cite{DEHP,KS} to
yield exact results for general $p$.  Furthermore, these calculations
provide the representations of the normalisation required for the
present work.  

These arise by finding the steady state weights $f(\Cee)$ for each
configuration $\Cee$ that satisfy the stationarity condition
\begin{equation}
\label{steadystate}
\sum_{\Cee^\prime \ne \Cee} \left[ f(\Cee^\prime) W(\Cee^\prime \to
\Cee) - f(\Cee) W(\Cee \to \Cee^\prime) \right] = 0
\end{equation}
in which $W(\Cee \to \Cee^\prime)$ is the probability of making the
transition from configuration $\Cee$ to $\Cee^\prime$ in a single
timestep.  A normalisation is then obtained through
\begin{equation} \label{Zgen}
Z = \sum_{\Cee} f(\Cee) \;.
\end{equation}
As remarked in the introduction, the weights are defined only up to an
overall scale.  In the calculations of \cite{Ev1}, a reduced form of the
full expression (\ref{trees}) was given in terms of the quantities
\begin{equation}
\fl
a_{n,r} = \sum_{t=0}^{n-r} \left[
 \left( \begin{array}{c} n\\r+t \end{array} \right)
 \left( \begin{array}{c} n-r-1\\t \end{array} \right)
 -
 \left( \begin{array}{c} n+1\\r+t+1 \end{array} \right)
 \left( \begin{array}{c} n-r-2\\t-1 \end{array} \right)
 \right] (1-p)^t
\label{anr}
\end{equation}
along with the conventions 
$\left( \begin{array}{c} X\\0 \end{array} \right)=1$ 
 and
$\left( \begin{array}{c} X\\-1 \end{array} \right)=0$
for any integer $X$.

With these definitions the normalisation $Z_L$ is
\begin{equation}
\label{zeeprime}
Z_L = Z^\prime_L + p Z^\prime_{L-1}
\;,
\end{equation}
where
\begin{equation}
Z^\prime_L = \sum_{r=0}^{L} a_{L,r} \frac{ 
(p(1-\beta)/\beta)^{r+1} - (p(1-\alpha)/\alpha)^{r+1} 
}{(p(1-\beta)/\beta) - (p(1-\alpha)/\alpha)}\;.
\label{Kexp}
\end{equation}
This should be contrasted with the simpler expression for the ASEP
with random sequential updates
\begin{equation}
\label{Zrs1}
\tilde Z_L = \sum_{r=1}^{L} \frac{r(2L-1-r)!}{L!(L-r)!}  \frac{
(1/{\beta})^{r+1} - (1/{\alpha})^{r+1} }{ (1/{\beta})-(1/{\alpha}) } 
\;,
\end{equation}
in which ${\alpha}$ and ${\beta}$ are transition \textit{rates} rather
than probabilities.

It is also possible to define {\it grand}-canonical partition
functions \cite{Ev1,Us1,Dep1} by summing over system sizes with
fugacity $z$ in both cases, so that ${\mathcal Z_p} (z) = \sum_L Z_L
z^L$ in the parallel case and $\tilde {\mathcal Z} (z) = \sum_L \tilde
Z_L z^L$ in the sequential one.  These definitions yield
\begin{equation}
\fl {\mathcal Z}_p (z) =\frac{ \alpha\beta (1+pz)
 \left[2(1-p)(\alpha\beta-p^2 z) -\alpha\beta b^2(1-p z) -\alpha\beta
 b^2\sqrt{(1+p z)^2-4z}\,\right]}
 {2p^4(1-\beta)(1-\alpha)(z-z_{hd})(z-z_{ld}) }\, \;,
 \label{Thfinal}
\end{equation}
where 
\begin{eqnarray}
b^2&=& \frac{p}{\alpha \beta}\left[ (1-p)-(1-\alpha)(1-\beta)\right]
\;,
\label{bsq}
\end{eqnarray}
and
\begin{eqnarray}
z_{ld} &=& {\alpha(p-\alpha) \over p^2(1-\alpha)}\;,
\label{aa1} \\
z_{hd} &=& {\beta(p-\beta) \over  p^2(1-\beta)}\;,
\label{aa2}
\end{eqnarray}
in the parallel case \cite{Ev1} and
\begin{equation}
\label{GCASEPz}
\tilde {\mathcal  Z} (z) = { 4 \alpha \beta \over (  
1 - 2 \alpha - \sqrt{ 1 - 4z }) ( 1 - 2 \beta -  \sqrt{ 1 - 4z }) }
\;,
\end{equation}
in the sequential case \cite{Us1,Heilmann,Dep1}.

The phase diagram can be extracted by considering the large $L$
behaviour of $Z_L$ and $\tilde Z_L$, either directly or from the
asymptotics of the grand-canonical generating functions. In the
parallel case using the latter approach \cite{Ev1} we can see that the
leading singularities in (\ref{Thfinal}) come from the poles at
$z_{ld}$, $z_{hd}$ in the low- and high-density phases respectively,
or from the square root singularity,
\begin{equation}
 z_{mc} = { 2 - p - 2 \sqrt{1 - p} \over p^2}\;,
\label{aa3}
\end{equation}
in the maximal current phase.
 
\begin{figure}
\begin{center}
\includegraphics[scale=0.25]{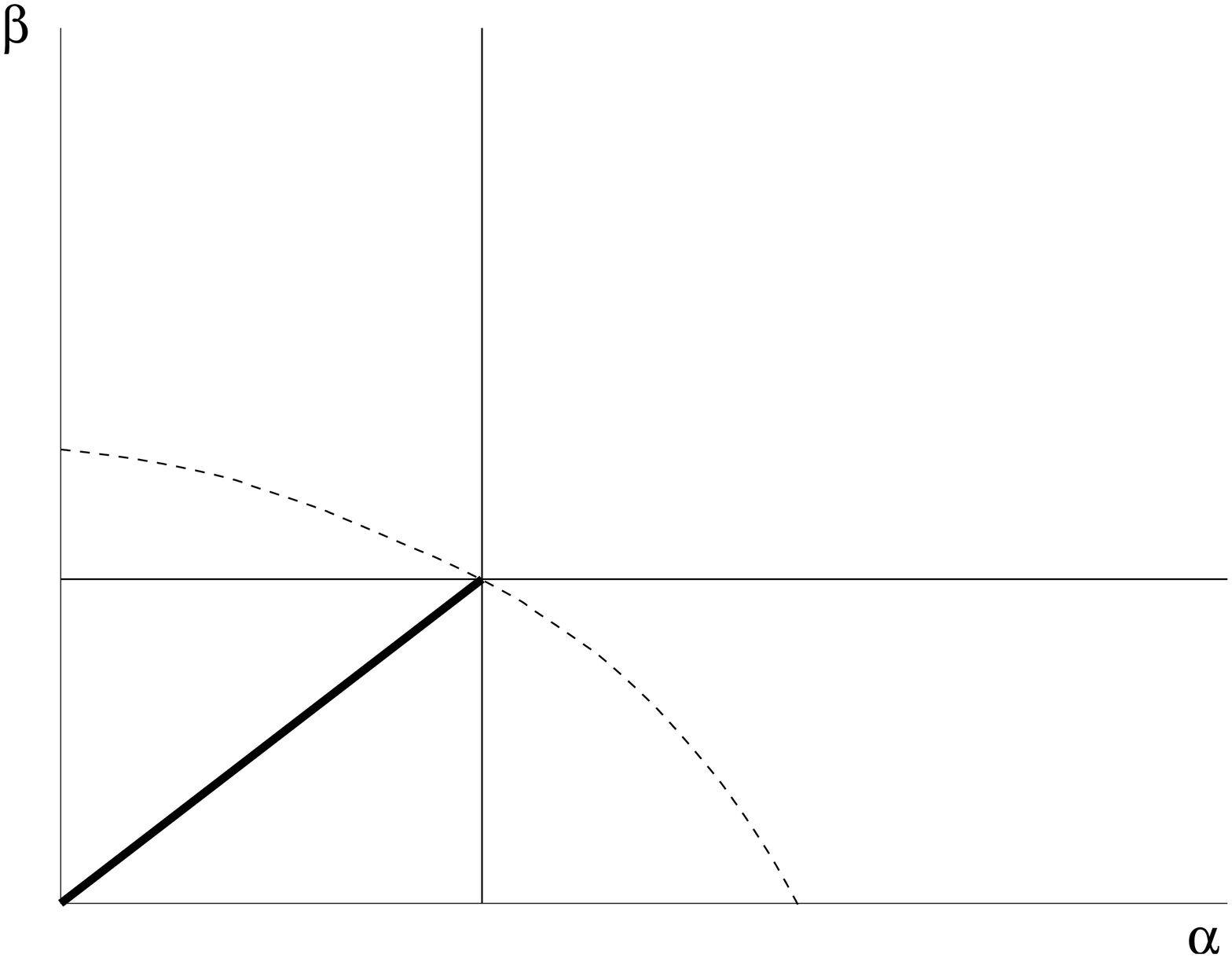}
\caption{\label{fig2}Phase diagram of the ASEP with parallel-update
dynamics.  The second-order lines are at $\alpha, \; \beta = 1 -
\sqrt{1 -p}$ and the first-order line in bold $\alpha = \beta$ runs
from the origin to meet these.  Also indicated as a dotted line is the
locus $ 1 - p = ( 1 - \alpha ) ( 1 - \beta)$ along which the mean
field solution is exact.}
\end{center}
\end{figure}

The phase diagrams for the random sequential and parallel updates have
similar topologies. That for the parallel case is shown in
Fig.~\ref{fig2}
where the main features to remark upon are the second-order transition
lines at $\alpha$ or $\beta = 1 - \sqrt{1 -p}$ and the first-order
line, depicted in bold, $\alpha = \beta < 1 - \sqrt{1-p}$. In
addition, the mean field solution is exact along the dotted locus $ 1
- p = ( 1 - \alpha ) ( 1 - \beta)$.  The region in which both $\alpha$
and $ \beta$ are greater than $1 - \sqrt{1 -p}$ is a maximal current
phase, that with $\alpha< 1 - \sqrt{1-p}$ and $\beta>\alpha$ is a
low-density phase, and that with $\beta< 1 - \sqrt{1-p} $ and $
\alpha>\beta$ is a high-density phase.  There are additional
transitions within these phases in which the bulk remains the same but
the boundary profiles change and these occur at $\alpha$ or $\beta = 1
- \sqrt{1 -p}$.  In a similar vein the phase diagram for the random
sequential updates ASEP shows a maximal current phase for $\alpha$ and
$ \beta >1/2$ separated from high- and low-density phases by second
order transition lines $\alpha=1/2, \, \beta>1/2$ and $\beta=1/2, \,
\alpha>1/2$.  There, the high- and low-density phases are separated by
a first-order line $\alpha=\beta \, <1/2$.

\section{Lee-Yang Zeros}
\label{zeros}

It has been observed that, for non-equilibrium systems, a possible
equivalent of the reduced free energy might be given by
\begin{equation}
f = - \lim_{L \to \infty} \frac{1}{L} Z_L.
\end{equation}
Since for parallel-update dynamics the singularities of
(\ref{Thfinal}) determine the asymptotic behaviour of the coefficients
$Z_L$, we can see that these will simply be given by
\begin{eqnarray}
\label{eq:fpdef}
f &=& \ln z_{mc} \;, \qquad{\rm for}\; \alpha, \; \beta> 1 -
\sqrt{1-p} \;, \nonumber \\
f &=& \ln z_{ld} \;, \qquad{\rm for} \;
\beta>\alpha, \; \; \alpha<1 - \sqrt{1-p} \;, \ \label{eq:fp<1}
\nonumber \\
f &=& \ln z_{hd} \;, \qquad{\rm for}\; \alpha > \beta ,
\; \; \beta<1 - \sqrt{1-p}\; .  \label{eq:fp>1}
\end{eqnarray}
For random sequential updates the singularities in (\ref{GCASEPz})
give
\begin{eqnarray}
\label{eq:fsdef}
\tilde f &=& \ln \frac{1}{4} \;, \qquad{\rm for}\; \alpha, \; \beta>
1/2\;, \nonumber \\
\tilde f &=& \ln \alpha ( 1 - \alpha)\;, \qquad{\rm for} \;
\beta>\alpha, \; \alpha<1/2 \; ,\ \label{eq:frsc<1} \nonumber \\
\tilde f &=& \ln \beta ( 1 - \beta )\;, \qquad{\rm for}\; \alpha >
\beta , \; \beta<1/2 \;, \label{eq:frsc>1}
\end{eqnarray}
where
$\tilde f$ is similarly defined as $ - \lim_{L \to \infty}  \tilde Z_L /L$.

It has been suggested that for the random sequential update ASEP and
other similar non-equilibrium models the Lee-Yang picture of
equilibrium phase transitions \cite{YL,LY,Fisher} might still apply
\cite{Arndt}.  The starting point of Lee and Yang's work was the
consideration of how the non-analyticity characteristic of a phase
transition appear given that the partition function is a polynomial on
finite lattices or graphs.  The form of the reduced free energy makes
it clear that any non-analyticities are associated with zeros of the
partition function when the appropriate fugacities (e.g. $y = \exp (
-2 h )$ for a spin model in field) are extended into the complex
plane.  In general, for a lattice with $L$ sites, the partition
function can be written as a polynomial in the fugacity,
\begin{equation}
Z_L = \sum_{r} D_{r} y^r \;,
\end{equation}
where the degree of the polynomial is proportional to $L$.  As the
polynomial may be completely expressed in terms of its zeros, $y_r$,
so too may the (reduced) finite-size free energy:
\begin{equation}
f_L (h) \sim  - {1 \over L } \ln \prod_{r} ( y - y_r (h))\;.
\label{blb}
\end{equation}
In the thermodynamic limit, $L \to \infty$, the zeros move in to pinch
the real axis at the transition point (or points) and the entire phase
structure of the model is determined by their limiting locus and
density. From (\ref{blb}), one has
\begin{equation}
f (h) \sim - \int_C d y \rho(y) \ln (y  - y(C) )\;,
\end{equation}
where $C$ represents some set of curves, or possibly regions, in the
complex $y$-plane on which the zeros have support and $\rho(y)$ is the
density of the zeros there.

The expressions for the reduced free energy for the ASEP allow a
direct determination of the locus of partition function zeros. In
general, the partition function zeros delineate the boundaries of
different phases since they give the loci along which the free energy
is singular, signalling a phase transition. In the Lee-Yang approach
one considers an extension to complex parameters, so the free energy,
$f$, becomes a complex quantity, $f_l$, which is the metastable free
energy per unit volume in the various phases given by $l=1,2..$, with
$\Re f_l = f$ characterising the free energy when phase $l$ is
stable. The loci of zeros are then determined by demanding $\Re f_i =
\Re f_j$ for adjacent phases $i,j$. Since we have $f = \ln
z_{mc,hd,ld}$ for the various phases of the parallel-update ASEP, we
can immediately see that the locus of zeros for the first-order
transition will be determined by
\begin{eqnarray}
\label{eq:lab}
| z_{hd} | = | z_{ld} |  \, , \quad \, \alpha, \beta < 1 - \sqrt{1-p}\; . 
\end{eqnarray}
Substituting in for $z_{hd}$, $z_{ld}$ from (\ref{aa1}) and
(\ref{aa2}), this gives
\begin{eqnarray}
\left|{\beta(p-\beta) \over  p^2(1-\beta)}\right| 
= \left|{\alpha(p-\alpha) \over p^2(1-\alpha)}\right| \; .
\end{eqnarray}
Similarly, the loci of zeros for the second order transitions are
determined by $| z_{hd}| = | z_{mc} |$ and $| z_{ld}| = | z_{mc} |$
or, from (\ref{aa1}), (\ref{aa2}) and (\ref{aa3})
\begin{eqnarray}
\label{eq:lc1}
\left|{\beta(p-\beta) \over  p^2(1-\beta)}\right| &=& 
\left|{ 2 - p - 2 \sqrt{1 - p} \over p^2} \right|\nonumber \\
\left|{\alpha(p-\alpha) \over p^2(1-\alpha)}\right| &=& 
\left|{ 2 - p - 2 \sqrt{1 - p} \over p^2} \right| \; .
\end{eqnarray}
These formulae are analogous to those observed in the case of the
first-order transition line in random sequential dynamics
\begin{equation}
| \alpha ( 1 - \alpha ) | = | \beta ( 1 - \beta ) | \; ,
\end{equation}
and for the second order lines
\begin{equation}
| \gamma ( 1 - \gamma ) | = 1/4 \; ,
\end{equation}
where $\gamma = \alpha, \beta$ depending on the region of the phase diagram.

For definiteness we now consider the zeros of the normalisation for
parallel-update dynamics given in (\ref{Zrs1}) with a convenient
rational value of $p= 16/25$.  This places the horizontal and vertical
second order transition lines of Fig.~2 at $\alpha$ and $ \beta = 1-
\sqrt{1-p} = 2/5$, respectively.  If we consider the former, we can
choose a fixed $\beta$ value, say $\beta = 3/5$, which will take us
across this line as $\alpha$ is varied.  Substituting $p=16/25,\,
\beta=3/5$ into the second of (\ref{eq:lc1}) gives the analytic locus
of zeros in the complex $\alpha$-plane which is plotted in Fig.~3.  We
have also used (\ref{Kexp}) to calculate the zeros numerically for
system sizes up to $L=250$, and these are plotted in Fig.~3, showing
quite good agreement with the limiting analytical results.
\begin{figure}
\begin{center}
\includegraphics[scale=0.5]{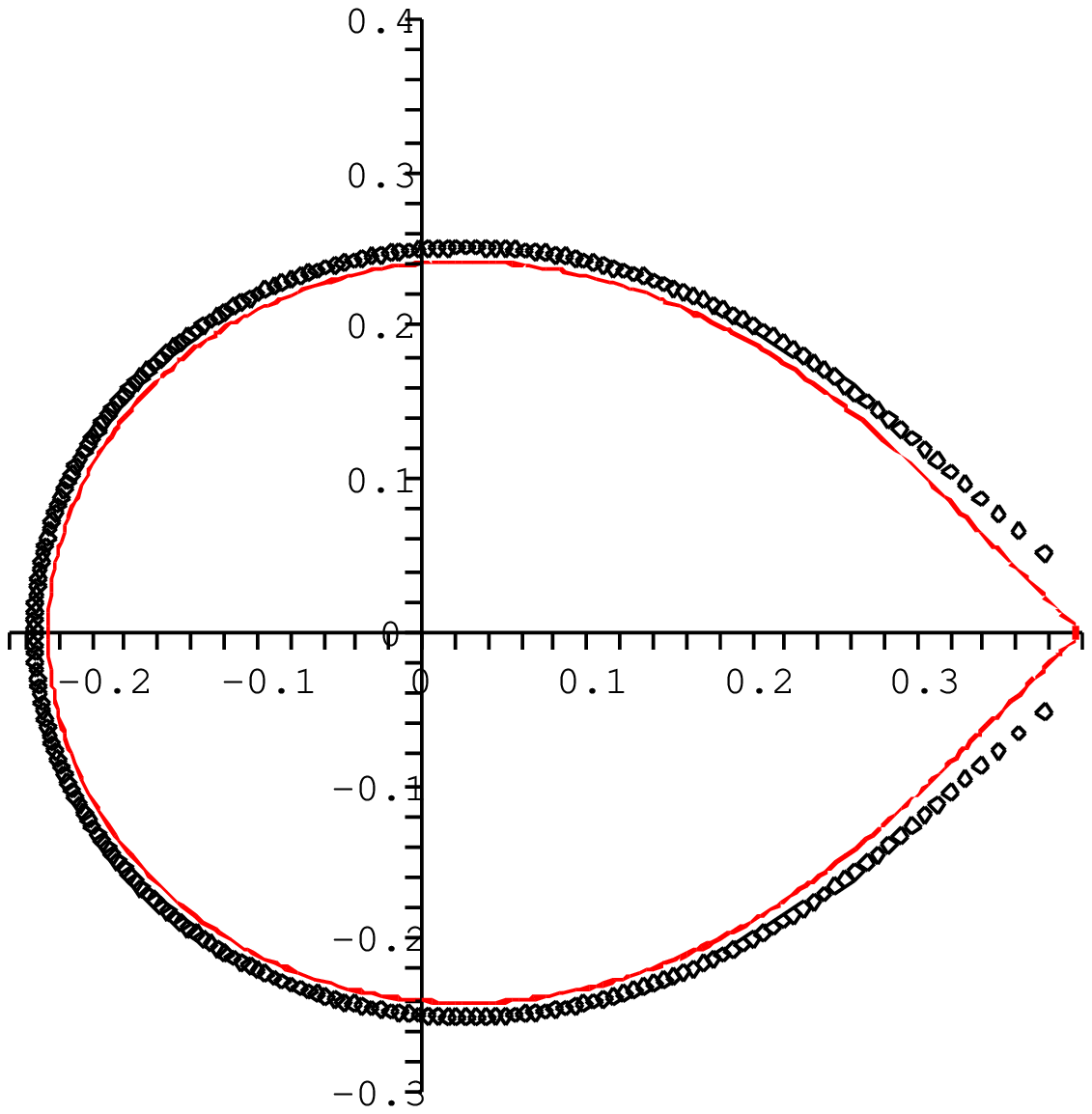}
\caption{\label{fig3} The locus of zeros in the complex $\alpha$-plane
for the parallel-update ASEP with $p=16/25,\, \beta=3/5$,
characteristic of a second-order transition. The diamonds show for
comparison numerically determined zeros for a system of size $L =
250$.}
\end{center}
\end{figure}
Similarly we can take a different value of $\beta$ which crosses the
first-order line, say $\beta=1/10$, and use (\ref{eq:lab}) determine
the analytic locus, which is shown in Fig.~4 along with the
numerically determined zeros for a system of size $L=250$.
\begin{figure}
\begin{center}
\scalebox{0.5}[0.5]{
\includegraphics{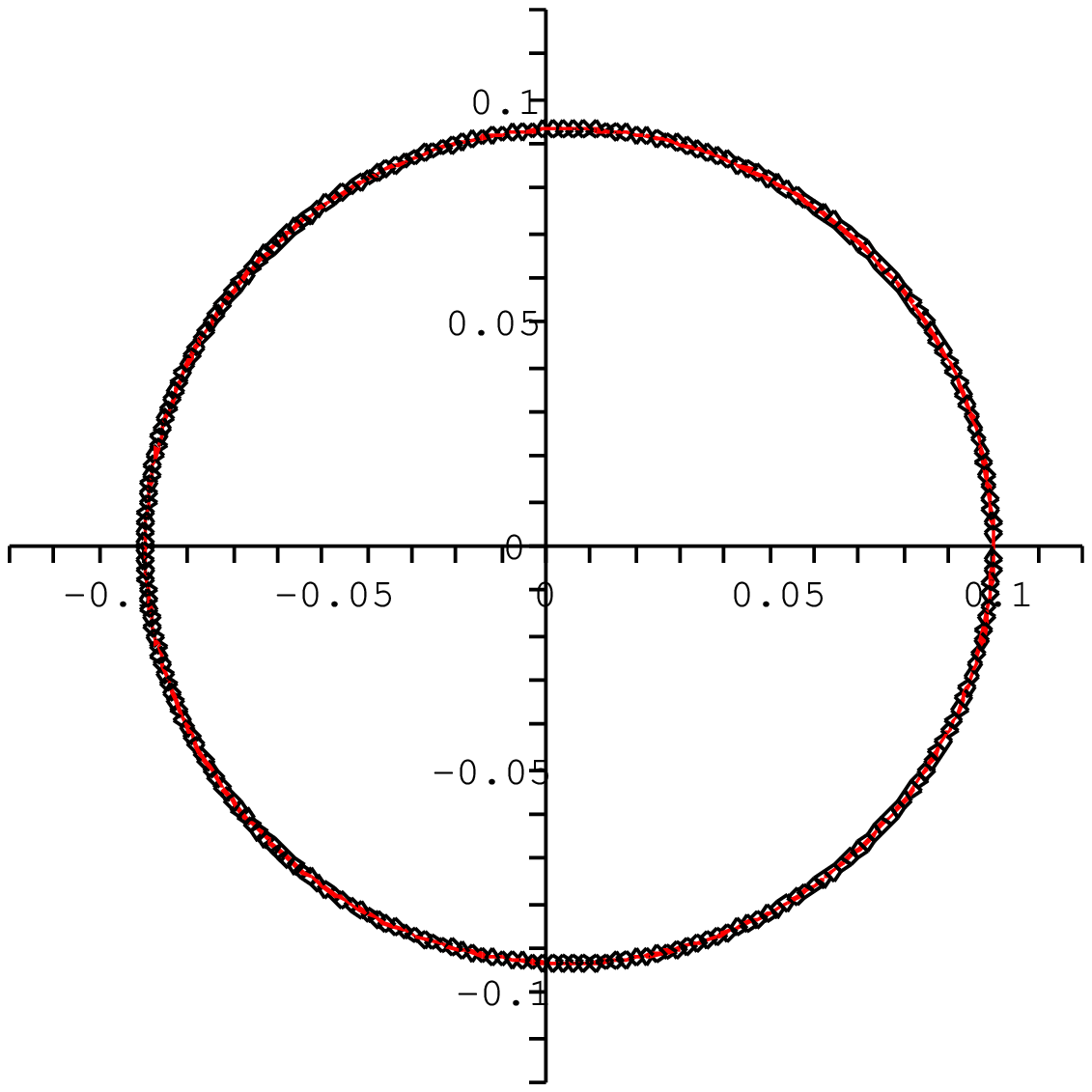}}
\caption{\label{fig4} The locus of zeros in the complex $\alpha$-plane
for the parallel-update ASEP with $p=16/25,\, \beta=1/10$
characteristic of a first-order transition. Again numerically obtained
zeros for a system of size $L = 250$ are superimposed.}
\end{center}
\end{figure}

Referring to the discussion of the partition function zeros for the
random sequential update ASEP in \cite{BE1}, we thus see the Lee-Yang
zeros technique still captures the nature of phase transitions with
parallel updates.  In Fig.~3 the zeros approach the phase transition
point at $\alpha = 2/5$ at a angle of $\pi / 4$ as might be expected
for a second order phase transition.  In the first-order case in
Fig.~4 the zeros approach the transition point at $\alpha = \beta =
1/10$ at an angle $\pi / 2$, which is characteristic of a first-order
transition.

\begin{figure}
\begin{center}
\scalebox{0.5}[0.5]{
\includegraphics{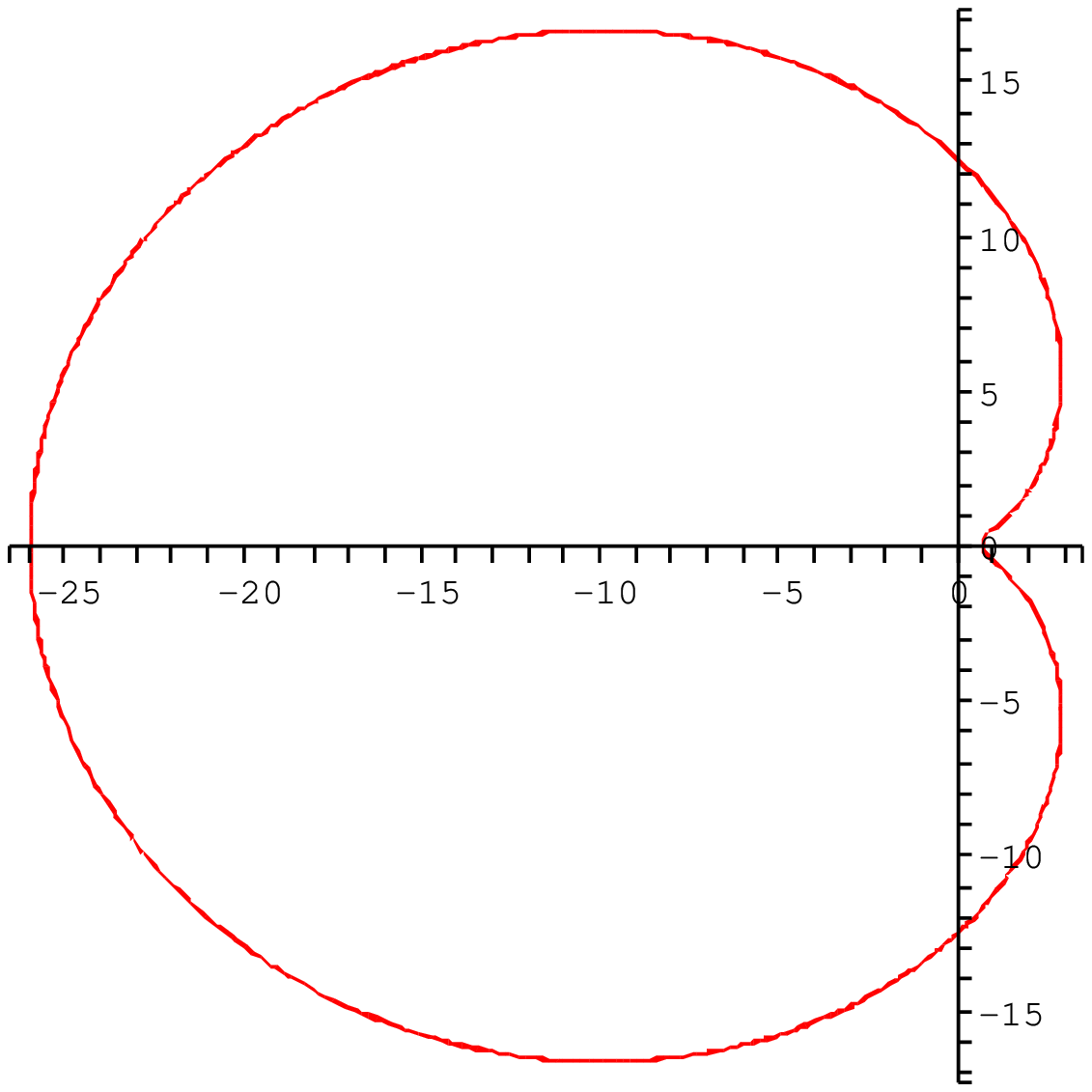}}
\caption{\label{fig4a} The locus of zeros in the complex $p$-plane for
the parallel-update ASEP with $\beta=2/5$. The transition point is
given by the cusp on the real axis at $1 - (1 - \beta)^2 = 16/25 =
0.64$.}
\end{center}
\end{figure}

It is also possible to look at zeros of the normalisation in the
complex $p$-plane for given $\alpha$ and $\beta$.  In this case one
would expect to see a transition at the larger of $1 - (1-\alpha)^2$
and $1-(1-\beta)^2$. The locus of zeros can be obtained by using, {\it
e.g.} $ |{\beta(p-\beta) / (p^2(1-\beta))}| = |{ ( 2 - p - 2 \sqrt{1 -
p} )/ p^2}|$ for fixed $\beta$ and complex $p$. The resulting cardioid
locus is shown in Fig.~\ref{fig4a}, with the transition appearing as
the cusp on the real axis.

\section{Thermodynamically Equivalent Equilibrium Models}
\label{teem}

The Lee-Yang analysis of the nonequilibrium normalisation for the
parallel-update ASEP in the preceding section suggests that the
transition probabilities defining the model are related to equilibrium
fugacities.  To establish this connection concretely, we seek
equilibrium models which are \textit{thermodynamically equivalent} to
the ASEP with parallel dynamics, which we define to occur when
\begin{enumerate}
\item the extensive part of the equilibrium free energy is identical
to that of the logarithm of the nonequilibrium normalisation specified
by equations (\ref{anr})--(\ref{Kexp}); and
\item the parameters $\alpha, \beta$ and $p$ are conjugate to
physical densities in the equilibrium model.
\end{enumerate}
A consequence of this definition is that (given the foregoing Lee-Yang
analysis of the nonequilibrium normalisation) such equilibrium models
will have the same thermodynamic phase behaviour.  An example of a
thermodynamic equivalence was recently provided for the
random-sequential limit $p \to 0$ by Brak \textit{et al.}\
\cite{Brak2,Brak1}.

Our method of choice in establishing such equivalences in the more
general case of parallel dynamics is to relate the grand-canonical
normalisation (\ref{Thfinal}) to generating functions for various
types of lattice walks.  In an earlier work \cite{Us1} we demonstrated
that in the limit $p\to 0$ this approach is technically simpler than
transforming the transfer matrices that build up the lattice-path
partition functions to the matrix product expressions used to solve
the ASEP, which was the method used in \cite{Brak2,Brak1}.

To this end we first note that (\ref{Thfinal}) can be expressed as
\begin{equation}
\mathcal{Z}_p (z) = { \alpha \beta ( 1 - x_{-}(z)^2/p ) \over (\alpha
- x_{-} (z) ) ( \beta - x_{-} (z) ) } \;,
\label{PGCASEPz}
\end{equation}
where 
\begin{equation}
x_{-} (z) = {p \over 2} \left( ( 1 + p z ) - \sqrt{ ( 1 +p z)^2 - 4 z
} \right)
\label{xminus}
\end{equation}
is one of the roots of the pole terms,
\begin{equation}
\label{zofx}
z = {x (p - x) \over p^2 (1- x)} \;,
\end{equation}
appearing in the grand-canonical partition function.

\begin{figure}
\begin{center}
\includegraphics[scale=0.75]{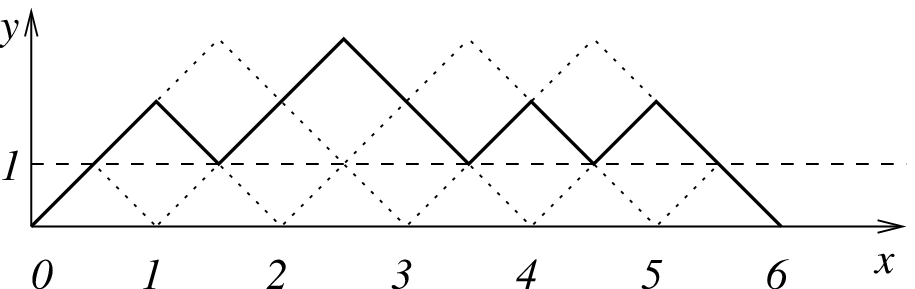}
\caption{\label{fig:excursion} An excursion on the rotated square
lattice.  Note that between the initial and final steps, the excursion
comprises sub-excursions away from the line $y=1$ and that the length
of the path along the $x$-axis is equal to the number of up-down
pairs.}
\end{center}
\end{figure}

Let us begin by briefly recapitulating the limit $p\to0$ for which the
equilibrium equivalence has been established \cite{Brak2,Us1,Brak1},
since the methods we employ will be of utility in the general case.
It is clear from (\ref{xminus}) that the function $\tilde{x}_{-} =
x_{-}/p$ remains finite in the limit $p\to 0$.  Therefore, to keep
(\ref{PGCASEPz}) finite we write
\begin{equation}
\mathcal{Z}_p (z) = 
{ [\alpha/p] [\beta/p] ( 1 - p [x_{-}(z)/p]^2 ) \over ([\alpha/p]  
- [x_{-}(z)/p] ) ( [\beta/p] - [x_{-}(z)/p] ) }
\;,
\end{equation}
which indicates the limit $p\to0$ must be taken with $\alpha/p =
\tilde{\alpha}$, $\beta/p=\tilde{\beta}$ as claimed in
Section~\ref{definition}.  Now note, from (\ref{zofx}), that the
function $\tilde{x}_{-}(z)$ satisfies
\begin{equation}
\label{zofxtilde}
z = \tilde{x}_{-} ( 1 - \tilde{x}_{-} )\;,
\end{equation}
in the limit $p \to 0$.  This expression implies that $\tilde{x}_{-}$
is the generating function of the number of \textit{excursions} on the
$45^\circ$ rotated square lattice, i.e.\ paths that touch the $x$-axis
only at the start and end.  The fugacity $z$ is conjugate to the
length of the path, measured in terms of the number of pairs of up-
and down-steps---see Fig.~\ref{fig:excursion}.

The generating function for excursions, $G_E(z)$, can be derived by
observing from Fig.~\ref{fig:excursion} that each comprises an initial
up-step and final down-step (which together contribute a fugacity $z$)
between which any non-negative number of excursions away from the line
$y=1$ can occur.  Therefore $G_E(z)$ can be expanded as
\begin{equation}
G_E(z) = z \left( 1 + G_E(z) + [G_E(z)]^2 + [G_E(z)]^3 + \cdots
\right) = \frac{z}{1-G_E(z)} \;.
\end{equation}
Comparison with (\ref{zofxtilde}) indicates that $\tilde{x} = G_E(x)$
as claimed.

\begin{figure}
\begin{center}
\includegraphics[scale=0.35]{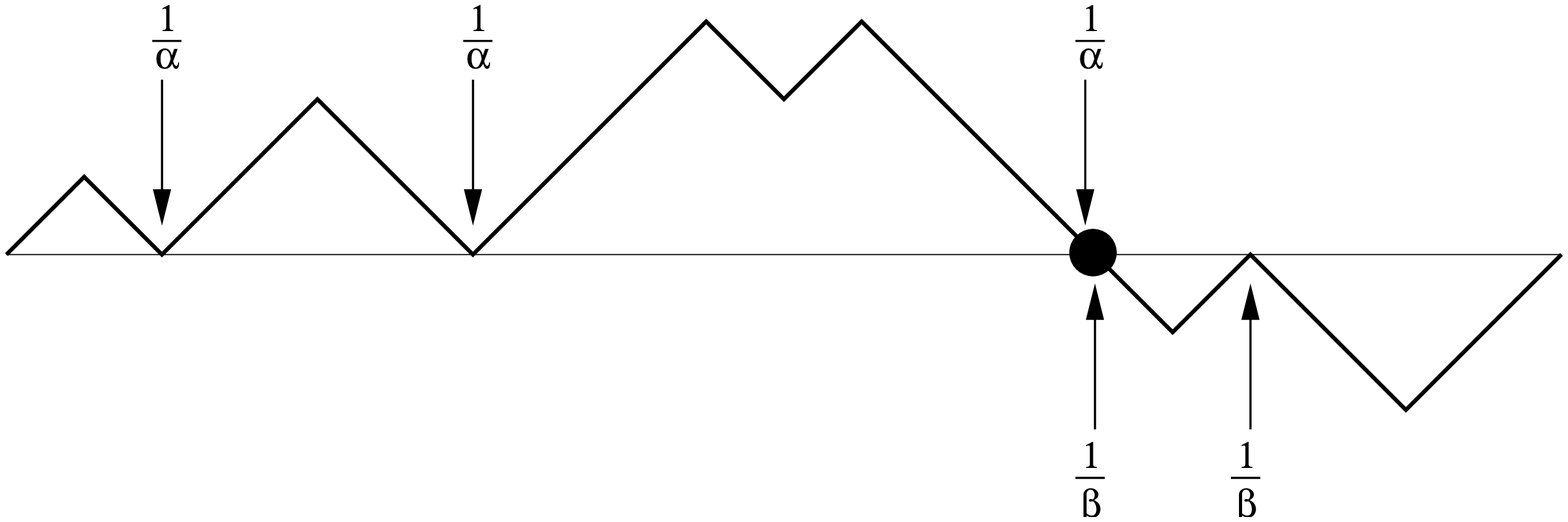}
\end{center}
\caption{\label{fig:OTW} A one-transit walk obtained by composing two
Dyck paths, one above the axis and one below. Each contact with the
$x$-axis from above, apart from the start, gives a factor $1/\alpha$
and each contact from below, apart from the end, gives a factor of
$1/\beta$.}
\end{figure}

To complete the path interpretation of (\ref{PGCASEPz}) in the
random-sequential limit, we recast in terms of the rescaled quantities
to obtain
\begin{equation}
\tilde{\mathcal{Z}}(z) = \lim_{p\to0} \tilde{\mathcal{Z}}_p(z) =
\frac{1}{(1 - \tilde{x}_{-}/\tilde{\alpha})
(1-\tilde{x}_{-}/\tilde{\beta})}\;,
\end{equation}
in agreement with \cite{Us1,Dep1}.  Expansion of the denominator
reveals that this expression describes an ensemble of paths with two
sets of excursions, one with a fugacity $1/\tilde{\alpha}$ per
excursion (or, equivalently, return to the $x$-axis) and the other
with fugacity $1/\tilde{\beta}$.  For convenience we take the first
set of excursions to go above the $x$-axis and the second below thus
defining a \textit{one-transit walk}, a realisation of which is shown
in Fig.~\ref{fig:OTW}.  In the mathematical literature walks that are
constrained to lie on or above the $x$-axis are often called
\textit{Dyck paths} (see e.g.~\cite{JvR}).  A one-transit walk is then
a composition of two Dyck paths.

We now use similar methods to establish equilibrium statistical
mechanical interpretations for the parallel-update ASEP with general
$p$.

\subsection{Walks on the triangular lattice}

\begin{figure}
\begin{center}
\includegraphics[scale=0.75]{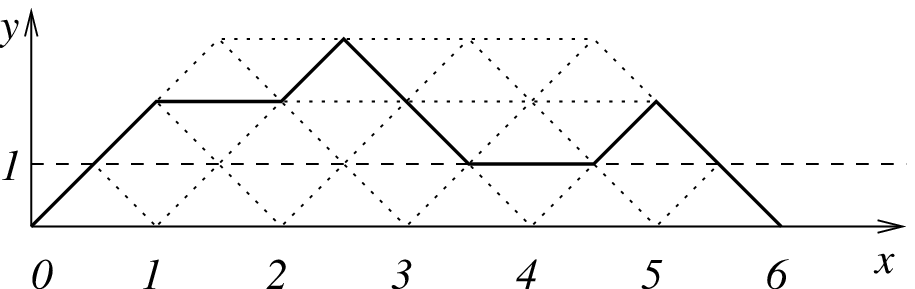}
\caption{\label{fig:triexcursion} An excursion on the triangular
square lattice.  In this case both horizontal segments and
sub-excursions occur on the line $y=1$.  The total length of the path
is the sum of the numbers of up-down pairs and horizontal segments.}
\end{center}
\end{figure}

Let us now consider excursions on a \textit{triangular} lattice, often
called {\it Motzkin} paths (such as that shown in
Fig.~\ref{fig:triexcursion}) in which each horizontal step contributes
a fugacity $u$ and an up-down pair $v$.  The generating function
$G_T(u,v)$ for these excursions can be derived as follows (but see
also, e.g.,~\cite{Brak3}).  As before, we note that each excursion
comprises an extremal up-down pair (contributing a factor $v$) and in
between there may be any combination of horizontal steps (each
contributing a factor $u$) and excursions (contributing $G_T(u,v)$),
including the null path.  That is
\begin{eqnarray}
G_T(u,v) &=& v \left( 1 + [u \!+\! G_T(u,v)] + [u \!+\! G_T(u,v)]^2 +
[u \!+\! G_T(u,v)]^3 + \ldots \right) \nonumber \\
&=& \frac{v}{1 - u - G_T(u,v)} \;.
\end{eqnarray}
Solving this equation one finds
\begin{equation}
G_T(u,v) = {1-u - \sqrt{ ( 1 - u)^2 - 4 v } \over 2}\;,
\label{GT}
\end{equation}
in which the negative root is taken to ensure all coefficients
appearing in the power-series expansion in $u$ and $v$ are positive.

A direct comparison of (\ref{GT}) with (\ref{xminus}) yields an
interpretation of the normalisation for the parallel-update ASEP in
terms of triangular-lattice walks with a fugacity $\sqrt{z}$ for each
up- or down-step and $-pz$ for each horizontal step.  This reads
\begin{equation}
\label{Ztriangle1}
\mathcal{Z}_p(z) = \frac{1 - p G_T(-pz,z)^2}{(1 - (p/\alpha)
G_T(-pz,z)) (1 - (p/\beta) G_T(-pz,z))} \;.
\end{equation}
Note that the power of $z$ in the generating function $G_T(-pz,z)$
measures the horizontal length of the path as shown in
Fig.~\ref{fig:triexcursion} since each up-down pair and horizontal
segment has the same length in the $x$-direction.

\begin{figure}
\begin{center}
\includegraphics[scale=0.35]{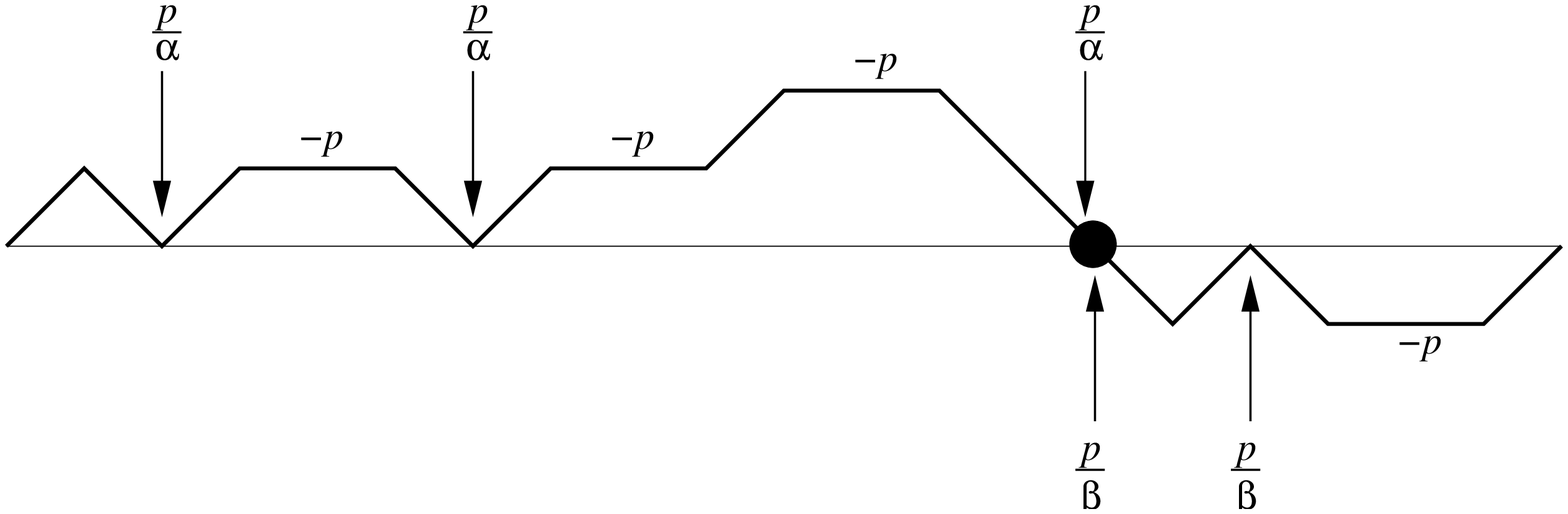}
\end{center}
\caption{\label{fig:triOTW} A one-transit walk on the triangular
lattice with contact weights for the axis.  The horizontal segments
have a weight $-pz$ whereas up-down pairs contribute a weight $z$, so
that the overall power of $z$ (not shown) is the horizontal length of
the path.  Meanwhile, there are contact fugacities of $p/\alpha$ from
above and $p/\beta$ from below.}
\end{figure}

We remark that only the denominator is relevant in terms of the
leading thermodynamic behaviour of the model.  One way to see this is
to note that the the factors in the numerator do not alter the
location or nature of the singularities in the complex plane.  More
precisely, if one calculates the canonical (fixed system-size) free
energy of the walk, it transpires that the numerator supplies only
\textit{subextensive} contributions (this point is demonstrated more
explicitly in Section \ref{thermo} below).  Hence we have established
a thermodynamic equivalence, as defined at the start of this section,
between the parallel-update ASEP and the one-transit walk on the
triangular lattice with fugacities $p/\alpha$ and $p/\beta$ associated
with contacts from above and below, a fugacity $z$ conjugate to the
path length and a negative fugacity $-p$ conjugate to the density of
horizontal segments as illustrated in Fig.~\ref{fig:triOTW}.

The origin of the negative fugacity can be understood from the
discussion of the Introduction.  Recall that one form of the
normalisation is the generating function of spanning in-trees on the
graph of configurations with edges weighted according to the
elementary transition probabilities, i.e.\ Eq.~(\ref{trees}).  Since
this model has discrete-time dynamics both the probabilities $\alpha,
\beta, p$ that particles move and their complements $1-\alpha,
1-\beta, 1-p$ that particles remain stationary appear as weights.  We
therefore do not require an expansion of the normalisation in powers
$p$ to have positive coefficients; however, we should expect to find a
series expansion in $p$ and $q=1-p$ with positive coefficients.

One can show this indeed the case for the function $G_T(-pz,z)$ and
thence the expansion of the denominator of (\ref{Ztriangle1}).
Expanding (\ref{GT}) and making the change of variable $p=1-q$ one
finds after some routine algebra that
\begin{equation}
G_T(-pz,z) = z \left( 1 + \sum_{n \ge 1} \sum_{m=1}^{n}  \frac{1}{n} 
{n \choose m} {n \choose m-1} z^n q^m \right)\;,
\label{triangleq}
\end{equation}
in which the coefficients of $z^n q^m$ are clearly positive. The
coefficients in (\ref{triangleq}) are the Narayana numbers
\cite{Narayana},
\begin{equation}
N (n,m) = \frac{1}{n} {n \choose m} {n \choose m-1}\;,
\end{equation}
which appear in many combinatorial contexts. Of particular note here
is the fact that they are known to count Dyck paths with $2 n$ steps
and $m$ peaks, so this is strongly suggestive that the parallel-update
ASEP also admits an interpretation involving standard Dyck paths with
weighted peaks. We shall see that this is indeed the case in the next
section.

\begin{figure}
\begin{center}
\includegraphics[scale=0.35]{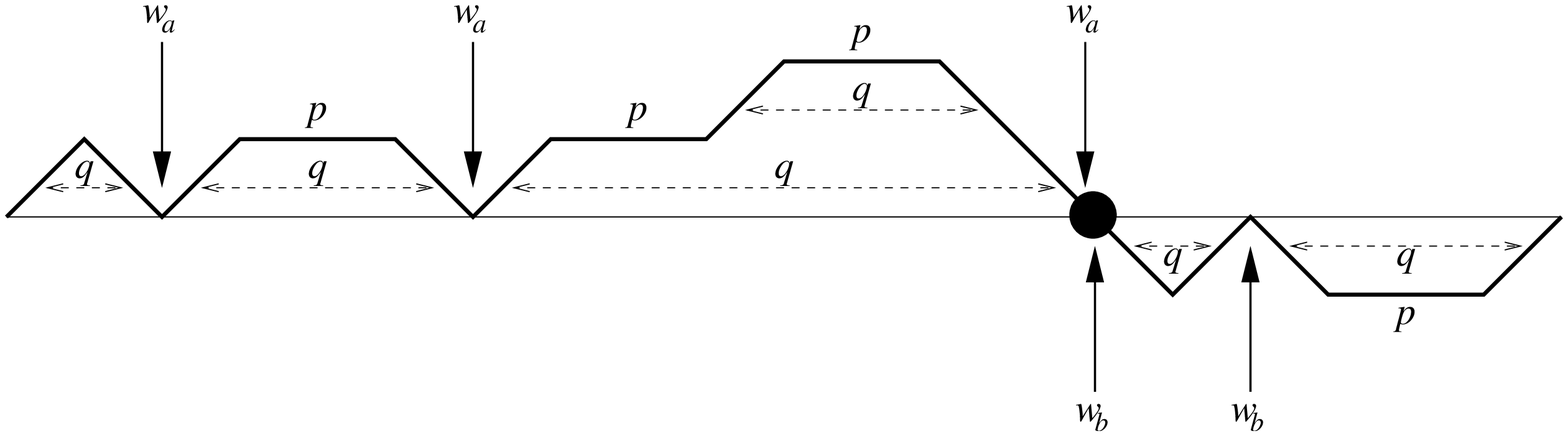}
\end{center}
\caption{\label{fig:triOTW2} An alternative one-transit walk on the
triangular lattice.  The horizontal segments have a weight $pz$
whereas up-down pairs contribute a weight $qz$.  Again the overall
(omitted) power of $z$ is the horizontal length of the path.  The
contact fugacities from above and below are modified to $w_a$ and
$w_b$ respectively.}
\end{figure}

It is also possible to find a triangular lattice path interpretation
of the normalisation (\ref{PGCASEPz}) in which the quantities $p$ and
$q$ appear more naturally as positive fugacities.  To do this, one
observes that, since $p+q=1$, (\ref{xminus}) can be rewritten as
\begin{equation}
\label{xmshifted}
x_{-}/p - pz = \frac{(1-pz) - \sqrt{(1-pz)^2 - 4qz}}{2} \;.
\end{equation}
Now the right-hand side of this expression is equal to $G_T(pz,qz)$,
i.e.\ the generating function of excursions on the triangular lattice
with a fugacity $q$ associated with each up-down pair and $p$ with
each horizontal step.  Expressing (\ref{Ztriangle1}) in terms of
$G_T(pz,qz)$ yields
\begin{equation}
\label{Ztriangle2}
\mathcal{Z}_p(z) = \frac{1 + (2p/q) G_T(pz,qz) - (p/q) G_T(pz,qz)^2}
{(1 - w_a G_T(pz,qz))(1 - w_b G_T(pz,qz))} \;,
\end{equation}
with
\begin{eqnarray}
\label{wa}
w_a &=& \frac{p(1-\alpha)}{q\alpha} \;, \\
\label{wb}
w_b &=& \frac{p(1-\beta)}{q\beta} \;\;.
\end{eqnarray}
As before, the terms in the numerator of (\ref{Ztriangle2}) are
thermodynamically unimportant.  Therefore the phase behaviour of the
parallel-update ASEP corresponds to that of a one-transit walk on the
triangular lattice with contact fugacities $w_a$ and $w_b$ from above
and below respectively, as shown in Fig.~\ref{fig:triOTW2}.

In both of the triangular lattice pictures one can understand the
limit $p\to0$ intuitively.  In this limit, paths with horizontal
segments are excluded from the partition function and one recovers the
one-transit walks on the rotated square lattice as discussed above.

\subsection{Weighted-peak walks on the rotated square lattice}

Another natural extension of the one-transit walk on the rotated
square lattice that arises in the case of parallel-update dynamics has
$q=1-p$ conjugate to the number of \textit{peaks} along the walk.  By
``peak'' we mean here a maximum above the $x$-axis, or a minimum
below.

\begin{figure}
\begin{center}
\includegraphics[scale=0.35]{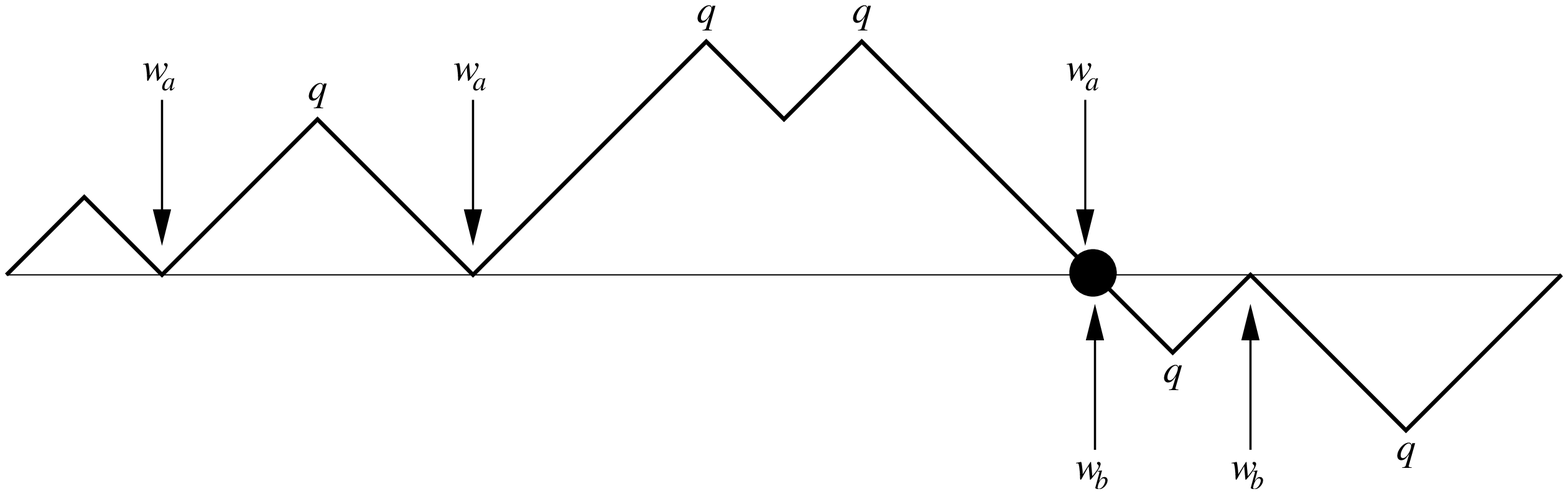}
\end{center}
\caption{\label{fig:peakOTW} A weighted-peak one-transit walk on the
rotated square lattice.  Here there is an additional weight $q$
associated with each maximum above the $x$-axis and minimum below.
The contact fugacities from above and below are the same as in
Fig.~\ref{fig:triOTW2}.}
\end{figure}

We begin, as before, by constructing the generating function
$G_P(z,q)$ of excursions on the rotated square lattice with the
addition of a fugacity, $q$, associated with each peak (a similar
derivation is given in \cite{Deutsch}).  In this instance one has
between the initial up-step and final down-step (which contributes a
factor $z$) either the null path (creating a peak with fugacity $q$)
or any positive number of excursions in which the number of peaks is
the sum of those from each excursion.  Thus we obtain the expansion
\begin{eqnarray}
G_P(z,q) &=& z \left( q + G_P(z,q) + [G_P(z,q)]^2 + [G_P(z,q)]^3 +
\cdots \right) \\
\label{GPfunctional}
&=& z \left( q + \frac{G_P(z,q)}{1-G_P(z,q)} \right) \;.
\end{eqnarray}
Solving this equation and discarding the unphysical root, we find
\begin{equation}
\label{GPexplicit}
G_P(z,q) = \frac{(1-pz) - \sqrt{(1-pz)^2 - 4qz}}{2} \;,
\end{equation}
which is none other than the right-hand side of (\ref{xmshifted}).
Hence we have immediately a representation of the parallel-update ASEP
normalisation
\begin{equation}
\label{Zpeak}
\mathcal{Z}_p(z) = \frac{1 + (2p/q) G_P(z,q) - (p/q) G_P(z,q)^2}
{(1 - w_a G_P(z,q))(1 - w_b G_P(z,q))}\;,
\end{equation}
whose denominator describes a one-transit walk on the rotated-square
lattice with each peak weighted by a fugacity $q$ and contact
fugacities $w_a$ and $w_b$ associated with contacts from above and
below and as given in the previous section as indicated in
Fig.~\ref{fig:peakOTW}.

Again, the numerator of this expression is thermodynamically
unimportant and we see that the limit $p\to0$ removes the weights
associated with the peaks.  This corresponds, as it should, to the
one-transit walks described at the start of this section in the
context of random-sequential dynamics.

\subsection{Other update dynamics}
\label{ordered}

Other update dynamics for the ASEP have been extensively investigated,
including sublattice parallel dynamics, and backward and forward
ordered sequential dynamics \cite{ScKo,ScPo}.  It was observed in
\cite{Ev1,dGN} that there was a close relation between sublattice
parallel dynamics, backward and forward ordered sequential dynamics
and the (totally) parallel dynamics we have discussed so far. For all
these the rules for making a hop are the same, but in the sublattice
parallel case all site pairs $i,~i+1$ with even $i$ are updated on
even time steps and all pairs with odd $i$ on odd time steps, whereas
in ordered sequential updates the sites are updated one at a time
starting at the left hand side and proceeding to the right (forward
ordered) or right to left (backward ordered).  The canonical
normalisation for the sublattice parallel ({\it sp}), backward ordered
sequential ({\it bos}) and forward ordered sequential dynamics ({\it
fos}) is given by the $Z_L$ of (\ref{zeeprime}), which in turn gives
the grand canonical normalisation ${\mathcal Z}_{sp,~bos,~fos}$
\begin{equation}
\fl
 {\mathcal Z}_{sp,~bos,~fos} (z) =\frac{
 \alpha\beta  \left[2(1-p)(\alpha\beta-p^2 z)
    -\alpha\beta b^2(1-p z)
    -\alpha\beta b^2\sqrt{(1+p z)^2-4z}\,\right]}
 {2p^4(1-\beta)(1-\alpha)(z-z_{hd})(z-z_{ld}) }\, \;,
 \label{Thfinal2}
\end{equation}
for all three updates \cite{Ev1}.

We can again rewrite (\ref{Thfinal2}) in a form which is recognizably
that of a generating function for one-transit walks.  Using $x_{-} (z)
= {p \over 2} \left( ( 1 + p z ) - \sqrt{ ( 1 +p z)^2 - 4 z } \right)$
we have
\begin{equation}
\mathcal{Z}_{sp,~bos,~fos} (z) = { \alpha \beta ( 1 - x_{-} (z)) \over
( \alpha - x_{-} (z) ) ( \beta - x_{-} (z) ) } \; ,
\label{SPGCASEPz}
\end{equation}
which should be compared to the expression for parallel updates in
(\ref{PGCASEPz}). With (\ref{SPGCASEPz}) at hand any of the walk
representations previously employed may be used to recast the
grand-canonical normalisation in the form of a one-transit walk
generating function. For instance, using the triangular lattice
generating function one can write
\begin{equation}
\mathcal{Z}_{sp,~bos,~fos} (z) = { ( 1 - p G_T(-pz,z)) \over ( 1 - ( p
/ \alpha ) \, G_T(-pz,z) ) ( 1 - ( p/ \beta ) \, G_T(-pz,z) ) } \; .
\label{spbosfos}
\end{equation}
Since the singularities in (\ref{spbosfos}) arise from the square root
in $G_T(-pz,z)$ and the poles in the denominator, which are identical
to the parallel case, we find the same phase diagram.
 
\section{Thermodynamics of One-Transit Walks}
\label{thermo}

To finish we briefly investigate the thermodynamics of one-transit
walks.  It is, in fact, possible to derive qualitatively the phase
diagram with only a little knowledge of the detailed structure of the
walks.  This provides a picture within which one can understand
clearly the robustness of the phase diagram of the ASEP under a
variety of updating schemes \cite{RSSS,ScKo,ScPo}.

Consider a walk of length $L$ that comprises $n_a$ excursions above
and $n_b$ excursions below the $x$-axis.  For added generality we
associate with each walk a number, $m$, that counts some property that
depends on the shape of the walk, but not on whether a particular
excursion lies above or below the axis.  The number of peaks is such a
property.  If there is an energy $\epsilon_a$ ($\epsilon_b$)
associated with each of the $n_a$ ($n_b$) contacts from above (below)
and an energy $\epsilon^\prime$ with the property counted by $m$, the
free energy of the walk is
\begin{equation}
\label{pathF}
F = n_a \epsilon_a + n_b \epsilon_b + m \epsilon^{\prime} - S(L, n_a,
n_b, m)\;,
\end{equation}
in which $S(L, n_a, n_b, m)$ is the entropy (logarithm of the number
of realisations) of paths with fixed $n_a, n_b$ and $m$.

The consequence of $m$ being independent of whether an excursion lies
above or below the axis is that
\begin{equation}
S(L, n_a, n_b, m) = \ln \Omega(L, n_a+n_b, m)\;,
\end{equation}
where $\Omega(L, n, m)$ is the number of walks of length $L$
comprising $n$ excursions and fixed $m$.  It then follows that if
(\ref{pathF}) can be minimised with respect to $n_a$, $n_b$ and $m$ in
such a way that $n=n_a+n_b$ is nonzero, this minimum is achieved by
all the excursions being on the side of the $x$-axis for which the
contact energy is smaller.  That is,
\begin{equation}
F = n \min\{\epsilon_a, \epsilon_b\} + m \epsilon^{\prime} - \ln
\Omega(L, n, m) \;.
\end{equation}
If $\epsilon_a = \epsilon_b$, the free energy is neutral to the
flipping of an excursion, and so along this coexistence line, the
crossing point will wander along the $x$-axis.  This is suggestive of
a first-order phase transition.

\begin{figure}[t]
\begin{center}
\includegraphics[scale=0.5]{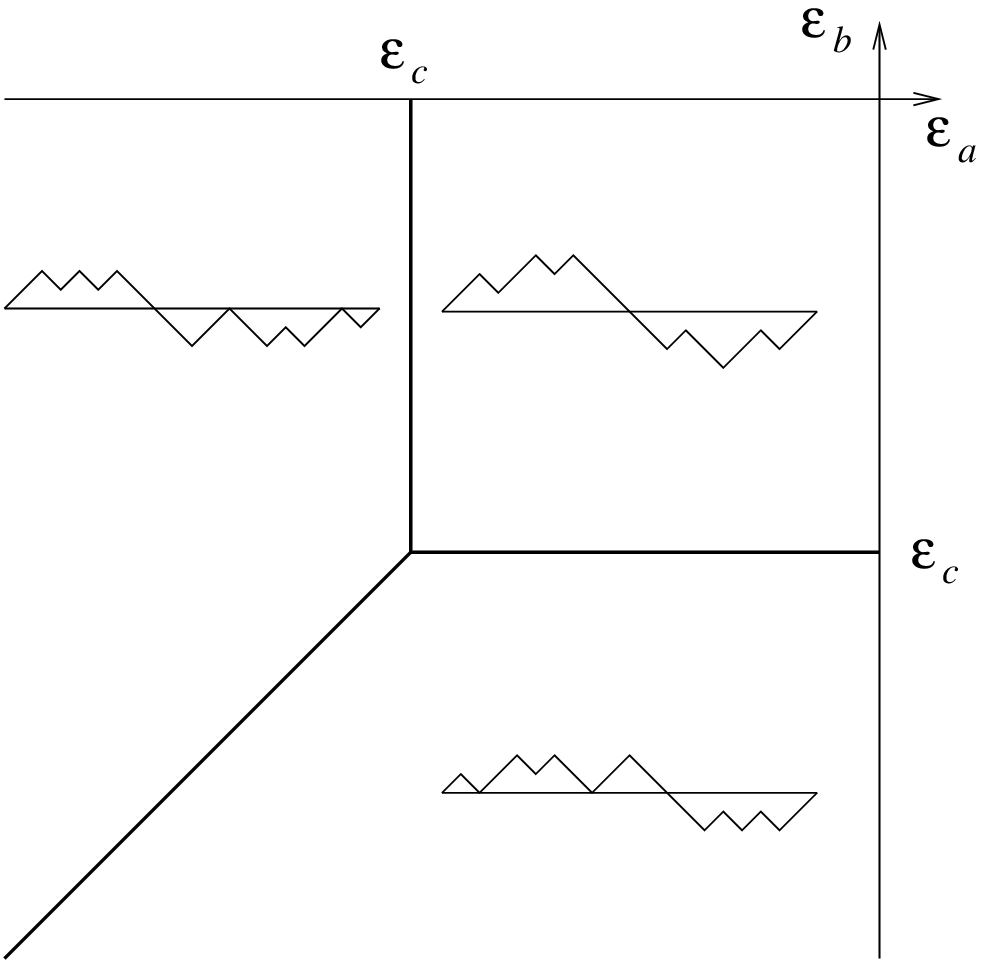}
\caption{\label{fig:OTWphase} Generic phase diagram of a one-transit
walk.  The density of contacts is nonzero on the side of the axis for
which the contact energy $\epsilon_{a,b}$ is lower.  The (negative)
critical energy $\epsilon_c$ at which the entropy of the walk causes
it to desorb from the axis depends on $\epsilon^\prime$ and how it is
coupled to the shape of the path.}
\end{center}
\end{figure}

It is intuitively clear that the number of possible walks increases as
the number of contacts decreases.  This implies that once the smaller
contact energy has increased above some critical value, the entropy
gained from avoiding the axis (i.e.\ becoming desorbed) is not offset
by any energy saving in contacting the axis.  Furthermore, since the
free energy depends only on the smaller of contact energies, the
critical value must be independent of the larger contact energy.  The
effect of $\epsilon^\prime$ will be to favour a particular shape of
path, and thus we would expect this to affect the critical contact
energy at which the path becomes desorbed.  These considerations give
rise to a phase diagram with the topology shown in
Fig.~\ref{fig:OTWphase} which is expected be common to a large class
of one-transit walks.

For concreteness, let us now consider the specific case of the
peak-weighted walk illustrated in Fig.~\ref{fig:peakOTW} which has $m$
as the number of peaks, $\epsilon_{a,b} = -\ln w_{a,b}$ and
$\epsilon^\prime = -\ln q$.  The number of walks of length $L$
comprising $n$ excursions and $m$ peaks is the coefficient of $z^L
q^m$ in $[G_P(z,q)]^n$.  This quantity can be obtained readily by
applying the Lagrange inversion formula to the functional relation
(\ref{GPfunctional}) for $G_P$ \cite{Deutsch,Wilf}.  This reveals that
for these walks, $\Omega(L, n, m)$ is the coefficient of $u^{L-n} q^m$
in
\[\frac{n}{L} \left( q + \frac{u}{1-u} \right)^L \;,\]
from which one quickly finds
\begin{equation}
\label{Omega}
\Omega(L, n, m) = \left\{ \begin{array}{ll}
\frac{n}{L} {L \choose m} {L-n-1 \choose m-n} & 0 < n \le m < L \\
1 & n = m = L \\
0 & \mbox{otherwise.}
\end{array} \right. 
\end{equation}
In the thermodynamic limit, $L\to\infty$, one finds the extensive part
of the entropy to be
\begin{eqnarray}
\fl s(\rho, \sigma) &=& \lim_{L\to\infty} \frac{1}{L} \ln \Omega(L,
\rho L, \sigma L) \nonumber\\
\fl&=& - \left[ \sigma \ln \sigma +
2(1-\sigma) \ln(1-\sigma) + (\sigma - \rho) \ln (\sigma-\rho) -
(1-\rho) \ln (1-\rho) \right]\;,
\end{eqnarray}
in which $\rho = n/L$ and $\sigma = m/L$ are the thermodynamic
densities of contacts and peaks.  Minimising the free energy per unit
length
\begin{equation}
f = \rho \min\{ \epsilon_a, \epsilon_b \} + \sigma \epsilon^{\prime} -
s(\rho, \sigma) \;,
\end{equation}
with respect to these densities one finds their equilibrium values are
\begin{eqnarray}
\rho^\ast &=& \frac{q(w-1)^2 - 1}{(w-1)(q[w-1]+1)}\;,\\
\sigma^\ast &=& \frac{q(w - 1)}{q(w-1)+1}\;,
\end{eqnarray}
for those values of $q$ and $w = \max\{ w_a, w_b \}$ where the
physical requirement $0 \le \rho^\ast \le \sigma^\ast \le 1$ holds.
One can check that the inequality $\rho^\ast \le \sigma^\ast \le 1$ is
automatically true whenever $\rho^\ast \ge 0$ is satisfied.  The
latter condition, however, is violated when $w_c < 1 + q^{-1/2}$.
Then the free energy is minimised at some point on the boundary of the
physical part of the $(\rho,\sigma)$-plane.  It is easily verified
that this point has $\rho^\ast=0$ (i.e.\ total desorption of the path)
and a peak density
\begin{equation}
\sigma^\ast = \frac{q^{1/2}}{1+q^{1/2}} \;.
\end{equation}
One then finds that in all phases the free energy takes the form
\begin{equation}
f = 2 \ln (1-\sigma^\ast) - \ln (1-\rho^\ast) \;.
\end{equation}
Using these results together with the definitions (\ref{wa}) and
(\ref{wb}), one recovers the expressions for the ASEP free energy
given in Section~\ref{zeros}.

This provides explicit confirmation that the peak-weighted one-transit
walk and the parallel-update ASEP are thermodynamically equivalent,
despite the fact the grand-canonical partition function of the latter
does not agree exactly with that for the ASEP (\ref{Zpeak}) which has
additional factors in its numerator.  The reason for the equivalence
is that these factors describe a small, fixed number of additional
excursions which contribute only to the subextensive part of the
entropy, and hence become irrelevant in the thermodynamic limit.

\section{Conclusions}

The Lee-Yang theory of partition function zeros has been shown to
apply to the ASEP with parallel-update dynamics, correctly pinpointing
the first- and second-order transition lines in the model in the
complex plane of transition probabilities. This fact suggests that
there is an interpretation of these transition rates as equilibrium
fugacities.  Indeed, we were able to make the correspondence concrete
by mapping the normalisation of the ASEP onto partition functions for
equilibrium path problems.  We found that for parallel updates there
were several possible path transcriptions, including two types of
adsorbing paths on a triangular lattice and a class of peak-weighted
paths on the rotated square lattice. In all cases the appropriate
scaling limit recovered the one-transit walk associated with the
random sequential update ASEP.

The benefit of finding such mappings is twofold.  First one sees quite
clearly why a Lee-Yang analysis of a nonequilibrium normalisation
should be appropriate in the complex plane of transition rates or
probabilities: under the mapping to an equilibrium model, one finds
that they \textit{are} fugacities.  Secondly, standard techniques from
equilibrium statistical mechanics can be used to learn about the
thermodynamics of nonequilibrium models.  For example, we showed that
the structure of the phase diagram for one-transit walks is
essentially unchanged when an additional fugacity associated with the
shape of the excursions is introduced.  In turn, this provides an
explanation as to why parallelising the dynamics of the ASEP does not
unduly change its phase behaviour.  This phenomenon can also be
understood in terms of an extremal current principle for driven
diffusive processes \cite{ScPo,Krug}.  This is suggestive that that
the equilibrium walk picture explored in the present work is
applicable to a wider range of processes to those studied so far.
Conversely, other nonequilibrium systems are known to be equivalent to
equilibrium models (such as the ``raise and peel'' interface model
\cite{Peel}) so one might expect Lee-Yang theory to apply to these
also.  Meanwhile, we note that the path equivalences have also been
utilised to calculate density fluctuations in the ASEP through a
suitable interpretation of the corresponding matrix-product
expressions \cite{DEL}.

In this work, we focussed on equilibrium models that are
\textit{thermodynamically equivalent} to the ASEP with parallel-update
dynamics.  At the start of Section~\ref{teem} we defined this
equivalence to exist when the extensive parts of the free energies
(defined for the nonequilibrium case as the logarithm of the
normalisation) coincide for the two models.  In practice, this meant
that the equilibrium surface models we described had partition
functions that differed from the normalisation of the ASEP.  This is
reflected in, e.g., nontrivial numerator factors in the
grand-canonical normalisations (\ref{Ztriangle1}), (\ref{Ztriangle2})
and (\ref{Zpeak}).  These factors did not concern us here, since we
were interested only in the thermodynamic phase behaviour of the
models: if one wishes to consider \textit{finite-sized} systems, one
should take care to include them.  Nevertheless, it turns out that
even for finite-sized systems, the loci of zeros of the normalisation
obtained from an ``artificial'' grand-canonical generating function
constructed from the denominator alone are similar to those obtained
using the full expressions.

Thus far, our analyses (and related calculations in \cite{Brak2,DEL})
have relied on the prior existence of solutions for the nonequilibrium
steady states under consideration.  This is, of course, an
unsatisfactory state of affairs, and ideally we would like to be able
to establish the path equivalences \emph{directly} from the definition
of the microscopic process.  We suggest, admittedly speculatively, that
the extremely compact forms of the grand-canonical normalisations for
the ASEP suggest that such an enterprise may be possible (though no
small challenge).  In particular, such a study might yield a new way
to determine nonequilibrium steady-state distributions.  As a first
step towards this goal, it would be of interest to investigate other
nonequilibrium states for which thermodynamically equivalent
equilibrium models have not yet been derived.  Chief among these is
the partially asymmetric exclusion process (PASEP) \cite{Sas,BECE} in
which particles may hop to the left as well as to the right across the
lattice.  When the left and right hop rates are equal, there is an
additional transition to a phase in which there is a zero current in
the thermodynamic limit.  If one could understand this transition from
an equilibrium statistical mechanical point of view, it would give
further weight to the growing evidence that phase transitions in
equilibrium and nonequilibrium steady states are two sides of same
coin.


\section{Acknowledgements}

WJ and DAJ were partially supported by EC network grant {\it
HPRN-CT-1999-00161}. RAB was supported by an EPSRC Fellowship under
grant {\it GR/R44768}.  We thank Martin Evans and Gunter Sch\"utz for
helpful comments.

\vspace{1cm}



\end{document}